\documentclass[a4paper,12pt]{article}

\usepackage{url,hyperref,lineno,microtype,subcaption}
\usepackage[onehalfspacing]{setspace}
\RequirePackage{inputenc,crop,graphicx,amsmath,array,color,amssymb,
flushend,stfloats,amsthm,chngpage,times,datetime,epstopdf, geometry} 
\usepackage{geometry}
\usepackage[english]{babel}
\usepackage{amsmath,amsfonts,amssymb,color,amsthm}   
\usepackage{pdfpages}
\usepackage{hyperref}
\usepackage{natbib}
\usepackage{rotfloat}
\usepackage[boxruled]{algorithm2e}
\usepackage[svgnames]{xcolor}

\newcommand{\newnnn}[1]{#1}
\newcommand{\new}[1]{#1}
\newcommand{\newn}[1]{#1}
\newcommand{\newnn}[1]{#1}

\begin{document}
\onecolumn

\title{\new{A flexible smoother adapted to censored data with outliers and its application to SARS-Co\newn{V}-2 monitoring in wastewater}}
\date{\vspace{-5ex}}

\author{Marie Courbariaux\,$^{1,*}$ \and Nicolas Cluzel\,$^{1}$ \and Siyun Wang\,$^{1}$ \and Vincent Maréchal\,$^{2}$ \and Laurent Moulin\,$^{3}$ \and Sébastien Wurtzer\,$^{3}$ \and \emph{Obépine} consortium \and Jean-Marie Mouchel\,$^{4}$ \and Yvon Maday\,$^{5}$ \and Grégory Nuel\,$^{1,6,*}$}


\maketitle
\footnotesize{\noindent \centering
$^{1}$Sorbonne Universit\'e, Maison des Mod\'elisations Ing\'enieries et Technologies (SUMMIT), 75005 Paris, France\\
$^{2}$Sorbonne Université, INSERM, Centre de Recherche Saint-Antoine, F-75012, Paris, France\\
$^{3}$Eau de Paris, Département de Recherche, Développement et Qualité de l’Eau, 33 avenue Jean Jaurès, F-94200 Ivry sur Seine, France\\
$^{4}$Sorbonne Université, CNRS, EPHE, UMR 7619 Metis, e-LTER Zone Atelier Seine, F-75005 Paris, France\\
$^{5}$Sorbonne Université, CNRS, Université de Paris, Laboratoire Jacques-Louis Lions (LJLL), F-75005 Paris, France\\
$^{6}$Stochastics and Biology Group, Probability and Statistics (LPSM, CNRS 8001), Sorbonne University, Campus Pierre et Marie Curie, 4 Place Jussieu, 75005, Paris, France\\
$^{*}$ corresponding authors: gregory.nuel@sorbonne-universite.fr, marie.courbariaux@sorbonne-universite.fr}

\normalsize

\begin{abstract}

A sentinel network, \emph{Obépine}, has been designed to monitor SARS-CoV-2 viral load in wastewaters arriving at 
wastewater treatment plants  \newnn{(WWTPs)} in France as an indirect macro-epidemiological parameter. 
The sources of uncertainty in such monitoring system are numerous and the concentration measurements it provides are left-censored and contain outliers, which biases the results of usual smoothing methods. 
Hence the need for  an adapted pre-processing in order to evaluate the real daily amount of virus arriving to each WWTP.\\ 
We propose a method based on an auto-regressive model adapted to censored data with outliers. Inference and prediction are produced via a discretised smoother which makes it a very flexible tool. This method is both validated on simulations and on real data from \emph{Obépine}.\\ 
The resulting smoothed signal shows a good correlation with other epidemiological indicators and \newnn{is currently used by \emph{Obépine} to provide an estimate of virus circulation over the watersheds corresponding to about $200$ WWTPs}.\\ %

\end{abstract}

\section{Introduction}

A sentinel network, \href{https://www.reseau-obepine.fr/}{\emph{Obépine (Observatoire épidémiologique dans les eaux usées)}}\footnote{\url{https://www.reseau-obepine.fr/}}, has been designed to monitor SARS-CoV-2 viral load 
in wastewaters arriving at \newn{about $200$}\footnote{in November 2021, representing more that one third of the French population} wastewater treatment plants (WWTPs) in France as an indirect macro-epidemiological parameter. 
This survey was initiated in March 2020 in WWTPs of the Greater Paris, during \newn{the} first epidemic wave in France \citep{wurtzer2020evaluation}. 
In this system, samples are currently typically taken twice a week, resulting in missing data on most days.
 
The sources of uncertainty in such monitoring system are numerous \citep{li2021uncertainties,arabzadeh2021data}. These include, for example: 
\begin{itemize}
\item sampling variance (although sampling at the WWTPs is integrated over 24 hours), \item variance from the Reverse Transcriptase (RT) and quantitative \newnn{or digital} Polymerase Chain Reaction (qPCR \newnn{or dPCR}) processes used to measure the concentration of SARS-CoV-2 RNA genes in the samples, 
\item uncertainties on other analytical steps in the labs such as virus concentration, genome extraction and the presence of inhibitors.
\end{itemize}
Using the raw viral load measurements right away to monitor the pandemic can therefore be misleading, as a large \newnn{variation} in the measured concentration can either be due to a real variation in virus concentration or to a quantification error. 
Therefore, these data need pre-processing in order to provide an \newn{accurate} estimate of the real daily amount of virus arriving to each WWTP and to evaluate the uncertainty on this estimate as also underlined by \cite{arabzadeh2021data}, who tested, for this purpose, $13$ concurrent smoothing methods on data from $4$ \newn{A}ustrian WWTPs. 

A solution to estimate the underlying true concentrations (from which the noisy measurements are derived) is to exploit the time dependence of the successive measurements.  
This temporal dependence is indeed due to this (never observed) underlying process, while the measurement noises from one time step to another are assumed to be independent. 
A natural way of exploiting this time dependence to denoise the signal is Kalman filtering \citep{kalman1960new,kalman1961new}. This is, for example, what is done for concentration monitoring by \cite{leleux2002applications} and it is one of the methods experimented by \cite{arabzadeh2021data}. 

Concentration measurements provided by the \emph{Obépine} network are left-censored because of detection and quantification limits in \newnn{(RT-)}qPCR  \citep[a problem pointed out by][for instance]{luo2013modelling}. \newnnn{This drawback, that is not taken into account in the work of \cite{arabzadeh2021data},} 
 results in a non-linear dependence between the measurements and the underlying process which cannot be handled by the classic Kalman filter. 
Such censoring obstacle to Kalman filtering can also happen in many other fields \citep[such as visual tracking, due to camera frame censoring, see][for instance]{miller2012kalman}. 
The Ensemble Kalman (EKF) permits to handle this problem by local 
linearization, but this linearization is sometimes not accurate enough 
which can lead to large errors \citep{wan2000unscented}. 
Unscented Kalman Filter \cite[UKF,][]{julier1995new} offers a solution to this problem with a enhanced statistical linearization around the state estimates. 
The Particle Filter \cite[PF,][]{mackay1992bayesian} also permits to handle censored data, but with a high computational cost. Finally, the Tobit Kalman filter \citep{allik2015tobit} is designed to handle censoring and has a reduced computational cost compared to PF  \citep{allik2015nonlinear}. In this approach, the censored data are replaced within an Expectation-Maximization algorithm by their conditional expectation with the current parameters.
However, we wish not only to predict in real time the value of the underlying process being measured, but also to reconstruct its past values and to estimate the error of the measurement system. To do so, a backward filter has to be computed and superimposed on the previously described (forward) filters, resulting in smoothers \citep{haykin2004kalman}. 

Data from the \emph{Obépine} network moreover contains numerous outliers (which may be caused by the vagaries of sampling or by a handling error, for example) that can bias a smoothing that would treat them as regular measurements. 
\cite{aravkin2017generalized} review adaptations of Kalman smoothing robust to such outliers (and also to abrupt changes of the underlying state and to switching linear regressions). This is mainly done by modifying the loss function to be minimized. However, those adaptations do not apply to the censoring scheme.

We here focus on the one-dimensional setting and propose a \new{Smoother adapted to Censored data with OUtliers (SCOU)} that answers both the censoring and the outliers detection problems through a discretization of the state space of the monitored quantities and more generally permits a very high degree of modelling flexibility.

The proposed model and its implementation are described Section~\ref{sec:method}.  
Section~\ref{sec:numeric} provides an illustration and validation of our approach using numerical simulations. 
Section~\ref{sec:appli} gives an example of utilization on data from the \emph{Obépine} network.

\section{The proposed method}
\label{sec:method}

\subsection{An auto-regressive model}
\label{sec:model}

Our method is based on the following state-space model: 
\begin{linenomath*}
\begin{eqnarray}
\label{eq:model}
X_t & = & \eta X_{t-1} + \delta + \sigma  \varepsilon_{X,t} \nonumber \\
O_t & \sim & \mathcal{B}(p) \nonumber \\
(Y^*_t|O_t=0) & = & X_t + \tau \varepsilon_{Y,t} \nonumber \\
(Y^*_t|O_t=1) & \sim & \mathcal{U}([a,b]) \\
Y_t & = & \max(Y^*_t,\ell) \nonumber\\
\begin{pmatrix}
\varepsilon_{X,t} \\
\varepsilon_{Y,t} \\
\end{pmatrix} & \overset{i.i.d}{\sim} & \mathcal{N}(0,I), \nonumber
\end{eqnarray}
\end{linenomath*}
where: 
\begin{itemize}
\item $t$ is the time index (ranging from $1$ to $n$). 
\item $X_t\in \mathbb{R}$ is the real quantity at time $t$ and   $X=(X_t)_{t\in\{1,...,n\}}$ is the vector of real quantities (to be recovered). 
\item $Y_t \in \mathbb{R}$ is the measurement at time $t$. $Y_t$ is generally only partially observed. We note $\mathcal{T}\subset\{1,...,n\}$ the set of $t$ at which $Y_t$ is observed. $Y=(Y_t)_{t\in\mathcal{T}}\newn{=Y_\mathcal{T}}$ is the vector of measurements. $Y^*$ is an accessory latent variable corresponding to a non-censored version of $Y$. $I$ is the identity matrix. 
\item $\eta\in\mathbb{R}$, $\delta\in\mathbb{R}$, $\sigma\in\mathbb{R}^+$\new{, $p\in [0,1] $}  and $\tau\in\mathbb{R}^+$ are parameters (to be estimated).
\item $\ell$ is the threshold below which censorship applies\footnote{In practice, $\ell$ can vary from one day to another, for instance if one works on quantities that correspond to the multiplication of concentrations (with a detection limit) by a fluctuating volume. This can be taken into account within our method with no additional cost.}.
\item  $O_t\in\{0,1\}$ is, for any $t \in \mathcal{T}$, the indicator variable of the event ``$Y^*_t$ is an outlier".  $O=(O_t)_{t\in\mathcal{T}}$. $\mathcal{B}(p)$ stands for the Bernouilli distribution of parameter $p$ and $\mathcal{U}([a,b])$ for the Uniform distribution on the interval $[a,b]$. $a$ and $b$ have to be chosen, they can for example correspond to quantiles (respectively very close to $0$ and very close to $1$) of the empirical marginal distribution of $Y$. 
\end{itemize}

\begin{figure}[h!]
\centering
\includegraphics[scale=0.15]{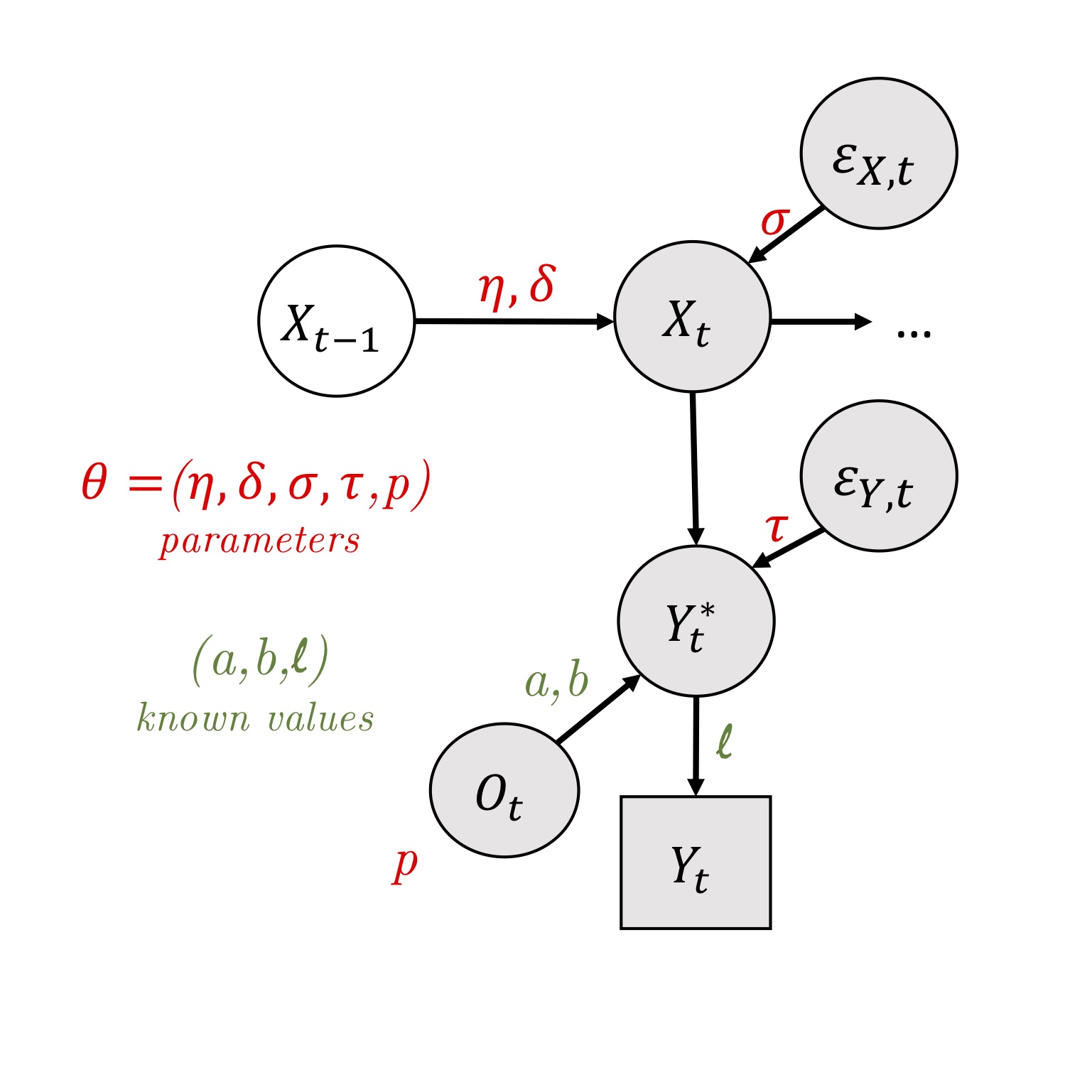}
\caption{\label{fig:diagr} Diagram of the proposed auto-regressive model. 
$(X_t)_{t\in\{1,...,n\}}$ is the underlying auto-regressive process (to be recovered), $(Y_t)_{t\in\mathcal{T}}$ is the measured quantity process, $Y^*_t$ a non-censored version of $Y_t$, $O_t$ is the indicator variable for the event "$Y^*_t$ is an outlier", $\varepsilon_{X,t}$ are tiny innovations and $\varepsilon_{Y,t}$ are the measurement errors.}
\end{figure}

Figure~\ref{fig:diagr} depicts this model. In this figure, 
the arrows represent the direct dependency relations between the variables. We can see the auto-regressive nature of the latent process $X$, while the $Y$ are independent conditionally to $X$ being known. 

\subsection{Inference and prediction - Smoothing and discretization}
\label{sec:inference}

We are interested in the distribution of 
$(X_t | \newn{Y_{\mathcal{T}}=y_{\mathcal{T}}})$ 
for every $t\in\{1,...,n\}$.

\subsubsection{The smoother}

This distribution can be computed from the forward ($F$) and backward ($B$) quantities, which  are defined as follows:
\begin{itemize}
\item $F$ is a (Kalman) Filter. For $t=1,\ldots,n$, 
\begin{linenomath*}
\begin{equation}
\label{eq:Fw}
F_t(x)=\mathbb{P}(X_t=x,Y_{\{1,...,t\}\cap\mathcal{T}}=y_{\{1,...,t\}\cap\mathcal{T}}).
\end{equation}
\end{linenomath*}
\item $B_n(x)=1$ (by convention). For $t=n-1,\ldots,1$, 
\begin{linenomath*}
\begin{equation}
\label{eq:Bk}
B_t(x)=\mathbb{P}(Y_{\{t+1,...,n\}\cap\mathcal{T}}=y_{\{t+1,...,n\}\cap\mathcal{T}}| X_t=x).
\end{equation}
\end{linenomath*}
\end{itemize}
For $t\in\{1,...,n\}$, $F_t(x)B_t(x) \propto \mathbb{P}(X_t=x|\newn{Y_{\mathcal{T}}= y_{\mathcal{T}}})$. 
In the \newn{strictly Gaussian} case, the Forward and Backward quantities can be computed recursively in closed formulas. Censorship and outliers unfortunately prevent the use of these closed formulas. 

\subsubsection{Discretization}
We  \newn{have} hence develop\newn{ed} a discrete version of those computations, which makes it easy to adapt to many possible extensions of the model (censorship but also outliers and possibly heteroscedasticity for instance). 
The  \newn{set of values that can be taken by $X$, approximated by $[a,b]$,} is discretized into $D$ values and becomes $\mathcal{X}$, with, for example, $\mathcal{X}=\{a,a+\Delta,a+2\Delta,...,b\}$ and $\Delta=\frac{b-a}{D-1}$. We then return to the framework of hidden Markov chains with $D$ states, the $D$ hidden states being the $D$ possible values for each $X$.

The transition matrix, $\pi$, is calculated as follows: 
\begin{linenomath*}
\begin{equation*}
\pi(x,x')=\mathbb{P}(X_{t+1}=x'|X_{t}=x)=\frac{f_\mathcal{N}(x';\eta x +\delta,\sigma^2)}{
\sum_{x''\in \mathcal{X}} f_\mathcal{N}(x'';\eta x +\delta,\sigma^2)},
\end{equation*}
\end{linenomath*}
where $f_\mathcal{N}(.;\mu,\sigma^2)$ is the Gaussian distribution function with mean $\mu$ and variance $\sigma^2$.

Let $e_t(x)=\mathbb{P}(Y_t=y_t|X_t=x)$. 
For all $x \in \mathcal{X}$, $e_t(x)=1$ if $t\notin \mathcal{T}$. If $t\in \mathcal{T}$: 
\begin{linenomath*}
\begin{equation*}
e_t(x)=(1-p) \left(\mathbb{I}_{\{y_t>\ell\}} f_\mathcal{N}(y_t;x,\tau^2) + \mathbb{I}_{\{y_t\leq \ell\}}F_\mathcal{N}(\ell;x,\tau^2)\right)
+p  \left(\new{\mathbb{I}_{\{y_t>\ell\}}  \frac{1}{b-a} + \mathbb{I}_{\{y_t\leq \ell\}}  \frac{\ell-a}{b-a}}\right), 
\end{equation*}
\end{linenomath*}
where $\mathbb{I}_{\{A=a\}}$ is the indicator variable for the event $\{A=a\}$ and $F_\mathcal{N}(.;\mu,\sigma^2)$ is the Gaussian cumulative distribution function with mean $\mu$ and variance $\sigma^2$. 

The Forward and Backward quantities, defined in Equations \ref{eq:Fw} and \ref{eq:Bk}, can then be calculated recursively from $\pi$ and $e$:
\begin{linenomath*}
\begin{itemize}
\item $F_1(x)=\newn{\frac{e_1(x)}{D}}$  and, for $t\in\{2,...,n\}$, $$F_t(x')=\sum_{x\in\mathcal{X}} F_{t-1}(x)\pi(x,x')e_t(x').$$  
\item $B_n(x)=1$ (by convention) and, for $t\in\{n,...,2\}$, $$B_{t-1}(x)=\sum_{x'\in\mathcal{X}} \pi(x,x')e_t(x')B_t(x').$$
\end{itemize}
\end{linenomath*}

Numerical tricks like logarithmic re-scaling are moreover used to calculate $e$, $\pi$ and the Forward and Backward quantities in order to avoid underflow and overflow problems.

\subsubsection{Learning} 
The parameters $\eta,\delta,\sigma,\tau$ and \new{$p$} maximizing the likelihood, 
which is given by $\sum_{x\in\mathcal{X}} F_n(x)$, 
\newn{could be obtained with the Expectation Maximization algorithm.
However, it is made challenging by the discretization and renormalization. This is why numerical optimization is preferred. The parameters are more precisely} obtained by the numerical optimization of Nelder and Mead (as implemented in the R function \texttt{optim}). 

\subsubsection{Prediction} 
For any $t$, the marginal distribution (mean, standard deviation) of $(X_t|\newn{Y_{\mathcal{T}}=y_{\mathcal{T}}})$, is given by $F_t(x)B_t(x) = \mathbb{P}(X_t=x,\newn{Y_{\mathcal{T}}=y_{\mathcal{T}}}) \propto \mathbb{P}(X_t=x|\newn{Y_{\mathcal{T}}=y_{\mathcal{T}}})$ (proof\footnote{ 
hint: decompose $\newn{(Y_{\mathcal{T}}=y_{\mathcal{T}})}=(Y_{\{1,...,t\}\cap\mathcal{T}}=y_{\{1,...,t\}\cap\mathcal{T}}, Y_{\{t+1,...,n\}\cap\mathcal{T}}=y_{\{t+1,...,n\}\cap\mathcal{T}})$, condition on $X_t=x$ and use the conditional independence.
} left to the reader) 
that is, for any $x\in\mathcal{X}$, the probability that $X_t$ (once discretized) takes the value $x$ conditionally to $Y_\mathcal{T}=y_\mathcal{T}$. 

\subsubsection{Simulations} 
We similarly show that:   
\begin{linenomath*}
\begin{equation*}
\mathbb{P}(X_{t-1}=x,X_t=x',\newn{Y_{\mathcal{T}}=y_{\mathcal{T}}})=F_{t-1}(x)\pi(x,x')e_t(x')B_t(x').
\end{equation*}
\end{linenomath*}
Thus, to simulate from $\mathbb{P}(X_{\{1,...,n\}}|\newn{Y_{\mathcal{T}}=y_{\mathcal{T}}})$, one can process sequentially, simulating $x_{1}$ with  $\mathbb{P}(X_1=x|\newn{Y_{\mathcal{T}}=y_{\mathcal{T}}})\propto F_1(x)B_1(x)$, then, for $t\in\{2,...,n]$: 
\begin{linenomath*}
\begin{equation*}
\mathbb{P}(X_t=x'|X_{t-1}=x,\newn{Y_{\mathcal{T}=y_{\mathcal{T}}}) = \frac{\pi(x,x')e_{t}(x')B_t(x')}{B_{t-1}(x)}}.
\end{equation*}
\end{linenomath*}

\subsubsection{Outliers detection} 
The probability for each observation $Y_t$ to be an outlier (knowing $Y_\mathcal{T}=y_\mathcal{T}$ and for the a priori probability of $p$) can be computed as follows:
\begin{linenomath*}
\begin{eqnarray}
\label{eq:outliers}
\mathbb{P}(O_t=1|\newn{Y_{\mathcal{T}}=y_{\mathcal{T}}})&=&
\frac{p \mathbb{P}(\newn{Y_{\mathcal{T}}=y_{\mathcal{T}}}|O_t=1)}{\mathbb{P}(\newn{Y_{\mathcal{T}}=y_{\mathcal{T}}})}\\ 
&=&\frac{p \sum_x\sum_{x'} F_{t-1}(x)\pi(x,{x'})B_t({x'}) \left(\new{\mathbb{I}_{\{y_t>\ell\}}  \frac{1}{b-a} + \mathbb{I}_{\{y_t\leq \ell\}}  \frac{\ell-a}{b-a}}\right)}{
\sum_x\sum_{x'} F_{t-1}(x)\pi(x,{x'})B_t({x'})e_{t}({x'})}. \nonumber 
\end{eqnarray}
\end{linenomath*}
An outlier is detected if $\mathbb{P}(O_t=1|\newn{Y_{\mathcal{T}}=y_{\mathcal{T}}})>h$, where $h$ has to be chosen according to the targeted false positive and false negative rates. 


\section{Numerical illustrations on artificial data}
\label{sec:numeric}

The proposed estimation is first evaluated on artificial data.  
The simulations are designed to study the ability of the algorithm to produce a good estimation of the parameters, to identify correctly the outliers and to adequately predict the conditional distribution of the underlying process.

\subsection{Data generation and protocol}
Artificial data are simulated according to the model~\eqref{eq:model} with $n=150$ time steps, $50\%$ of which are observed. 
We set $a$ and $b$ respectively to the quantiles $0.02\%$ and $99.98\%$ of the marginal distribution of $Y'$.
For the sake of consistency with the study presented in Section~\ref{sec:appli}, parameters are set so as to be realistic with regard to the \emph{Obépine} data:  $\sigma=0.3$, $\tau=0.6$, $\eta$ close to $1$ and $\delta$ close to $0$.

We realize five experiments: 
\begin{itemize}
\item with no censoring, no outlier, $\delta=0$, $\eta=1$, 
\begin{itemize}
\item (1) with $\Delta=0.02$, 
\item (2) with $\Delta=0.1$,
\item (3) with $\Delta=0.7$, 
\end{itemize}
\item (4) with $\Delta=0.1$, $\ell$ set for each simulated data such that $16\%$ of $Y'$ are censored (medium censoring level), an outlier rate of $p= \new{7}\%$, $\eta=0.99$ and $\delta=0.001$,
\item (5) same setting as the previous experiment but with $31\%$ of censored data (high censoring level).
\end{itemize}
Each experiment is replicated $100$ times.
For each simulation, we compare our approach\new{, SCOU,} to~:
\begin{itemize}
\item 
a 2-parameter Kalman Smoother implemented in the DLM R package \citep{dlm2010}. In the latter method, censoring and outliers are not taken into account and $\delta$ and $\eta$ are not estimated ($\delta$ is set to $0$ and $\eta$ is set to $1$). Those hypotheses correspond to the settings of the experiments (1), (2) and (3),  which aims at checking that the two methods give identical results if $\Delta$ is small enough.
\item \newnnn{the moving average smoother.} 
\item \newnnn{the LOESS \citep{atkeson1997locally} which consists in locally weighted polynomial regressions.} 
\end{itemize}

\new{As in the study by \cite{arabzadeh2021data}, the parameters of the two last methods (the number of days of the moving average smoother and the span parameter of the LOESS) are chosen by leave-one-out cross-validation so as to minimize the Root Mean Squared Error (RMSE) for the prediction of the left out observations.}  
When (left-)censoring occurs, considering the observed censored values would result in a poor performance of those smoothers in recovering the underlying auto-regressive signal. We compensate this by instead taking 
\new{the estimated mean for the corresponding right-truncated normal distribution in $\ell$ (estimated by assuming that the observations follow a left-censored normal distribution in $\ell$)} 
as the observed value for those smoothers, 
which results in improved performances in all cases.

\subsection{Results on artificial data}

\subsubsection{Impact of discretization}

\begin{figure}[h!]
\centering
\includegraphics[scale=0.2]{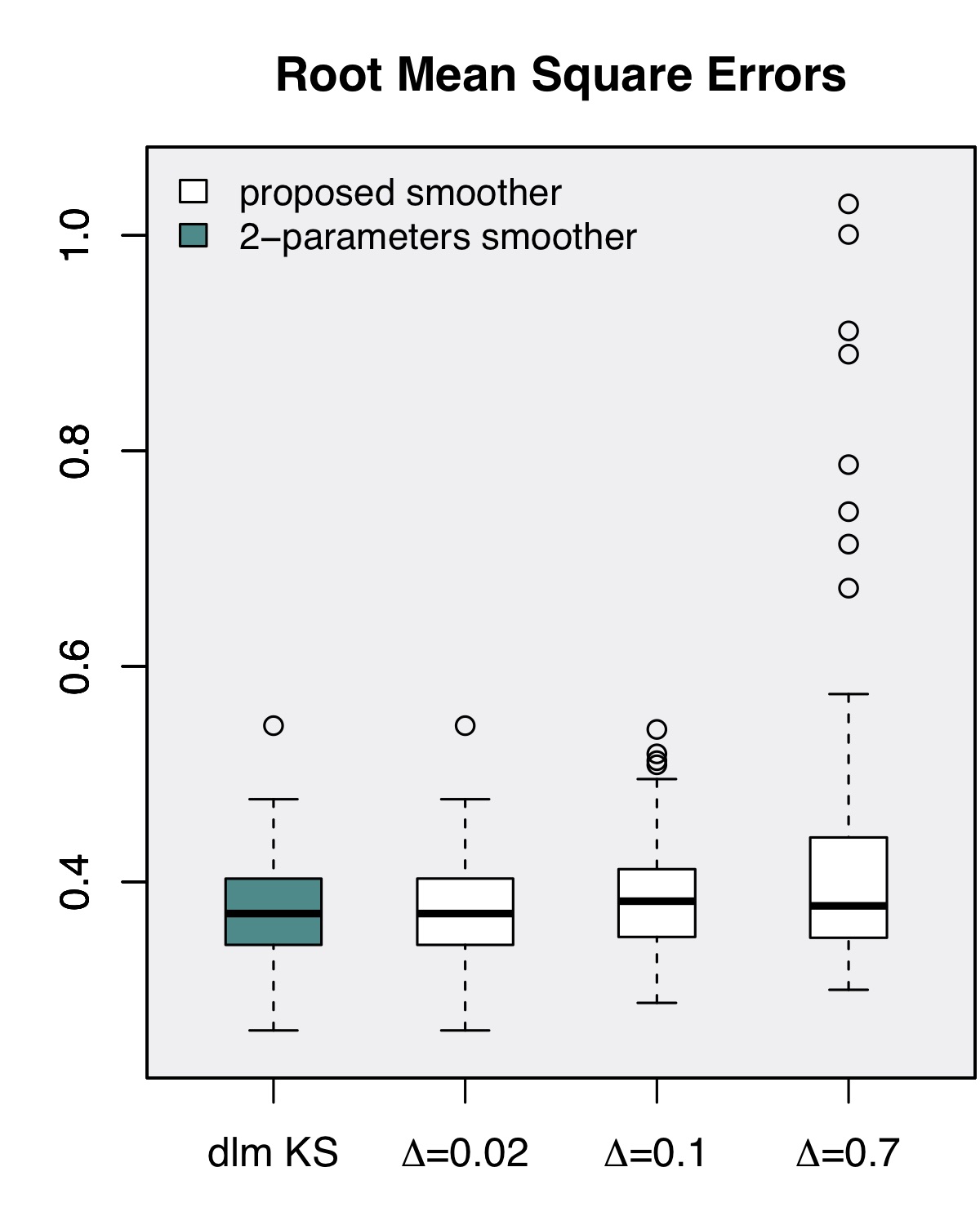}
\caption{\label{fig:DELTA} RMSE \newn{obtained for the prediction of the true underlying signal by the} two-parameter exact smoother and by our method on data sets simulated with no outliers, no censoring, $\delta=0$ and $\eta=1$ and with varying values of the discretization step, $\Delta$. \new{As expected, in this ideal setting, the 2-parameters Kalman smoother implemented in the DLM R package and our method give identical results for $\Delta=0.02$.}}
\end{figure}

The first, second and third experiments (no outliers, no censoring,  $\delta=0$ and $\eta=1$) show, as expected, that the 2-parameters Kalman smoother implemented in the DLM R package and our method give identical results for $\Delta=0.02$. 
The two methods show close performances for $\Delta=0.1$ and a substantial degradation when $\Delta=0.7$, as shown   Figure~\ref{fig:DELTA}, where the RMSE (Root Mean Square Error) for the prediction of $X$ is computed for each of the $100$ repetitions of each of the three experiments.

In what follows, we focus on the two other experiments (medium and high censoring levels, $p=\new{7}\%$, $\eta=0.99$, $\delta=0.001$ and $\Delta=0.1$).

\subsubsection{Illustration on one simulation example}

\begin{figure}[h!]
\centering
\includegraphics[scale=0.15]{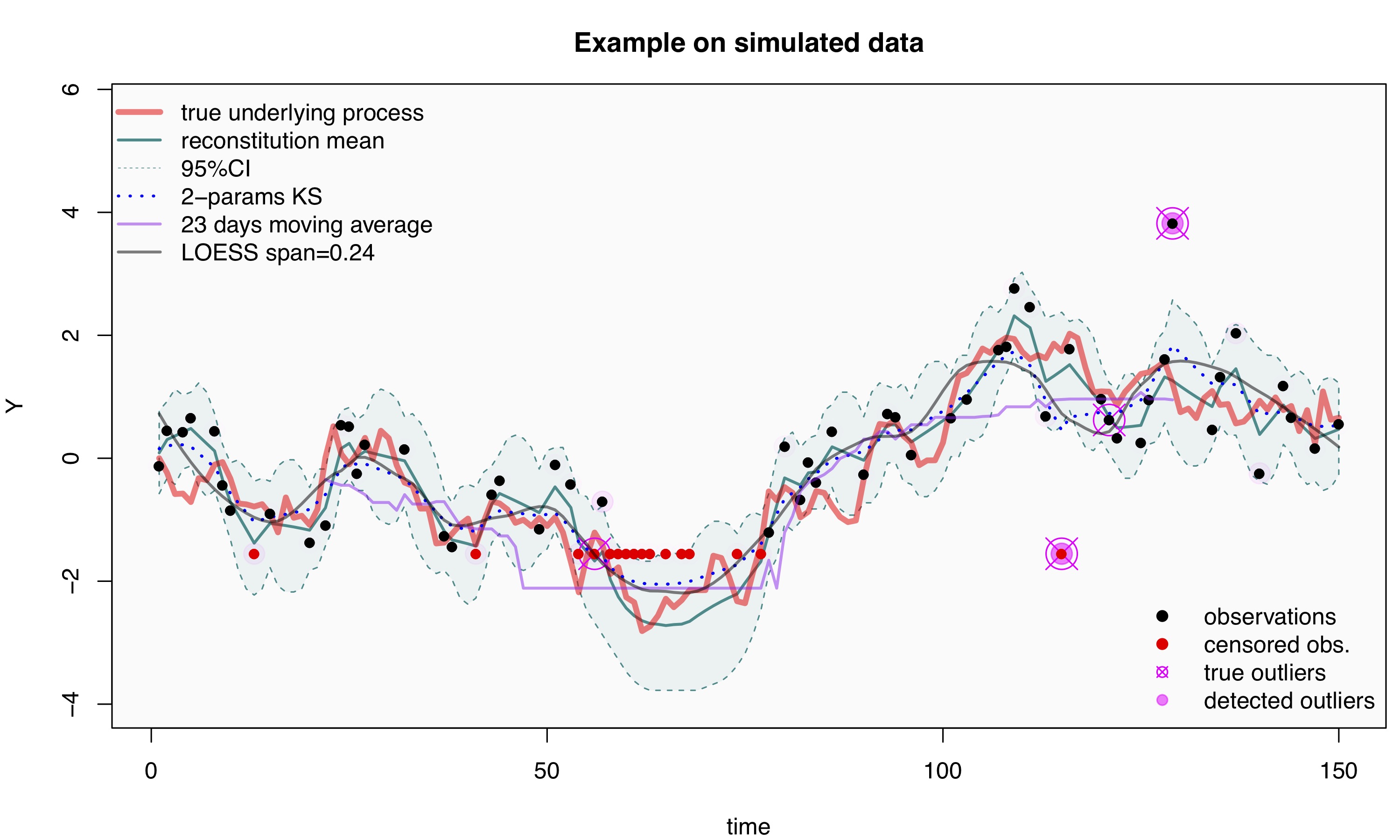}
\caption{\label{fig:datasim} Simulated data with $16\%$ of censored data and $p=\new{7}\%$ of outliers \newnnn{and results from the proposed (smoothing, outlier detection and prediction) method, the equivalent 2-parameter Kalman Smoother, the $23$-days moving average smoother and the LOESS with $0.24$ as span parameter}.}
\end{figure}

Figure~\ref{fig:datasim} illustrates on an example the results of our method on simulated data within the fourth experiment setting (medium censoring level). On this illustration, we can see that the learning process and the use of the smoother permit to finely predict the trajectory of the underlying process, $X$ (in red), from the observations $Y_\mathcal{T}=y_\mathcal{T}$ (represented by dots) while adequately taking into account the time interval during which the censoring applies. 
On this figure, the more pink-colored the points are, the higher the estimated probability of them being an outlier. For a detection threshold set lower than $h=\new{0.95}$, \new{two out of the four} simulated outliers (pink crosses and circles, two of them being censored) are identified. 
\newnnn{The $23$ days moving average and the LOESS (for span$=0.24$) smoothers\new{, which parameters are selected by leave-one-out cross-validation,} give rather good reconstitutions but fail in reconstructing the trajectory when censoring or outliers happen.}

\subsubsection{Parameters estimation}
\begin{figure}[h!]
\centering
   \includegraphics[scale=0.2]{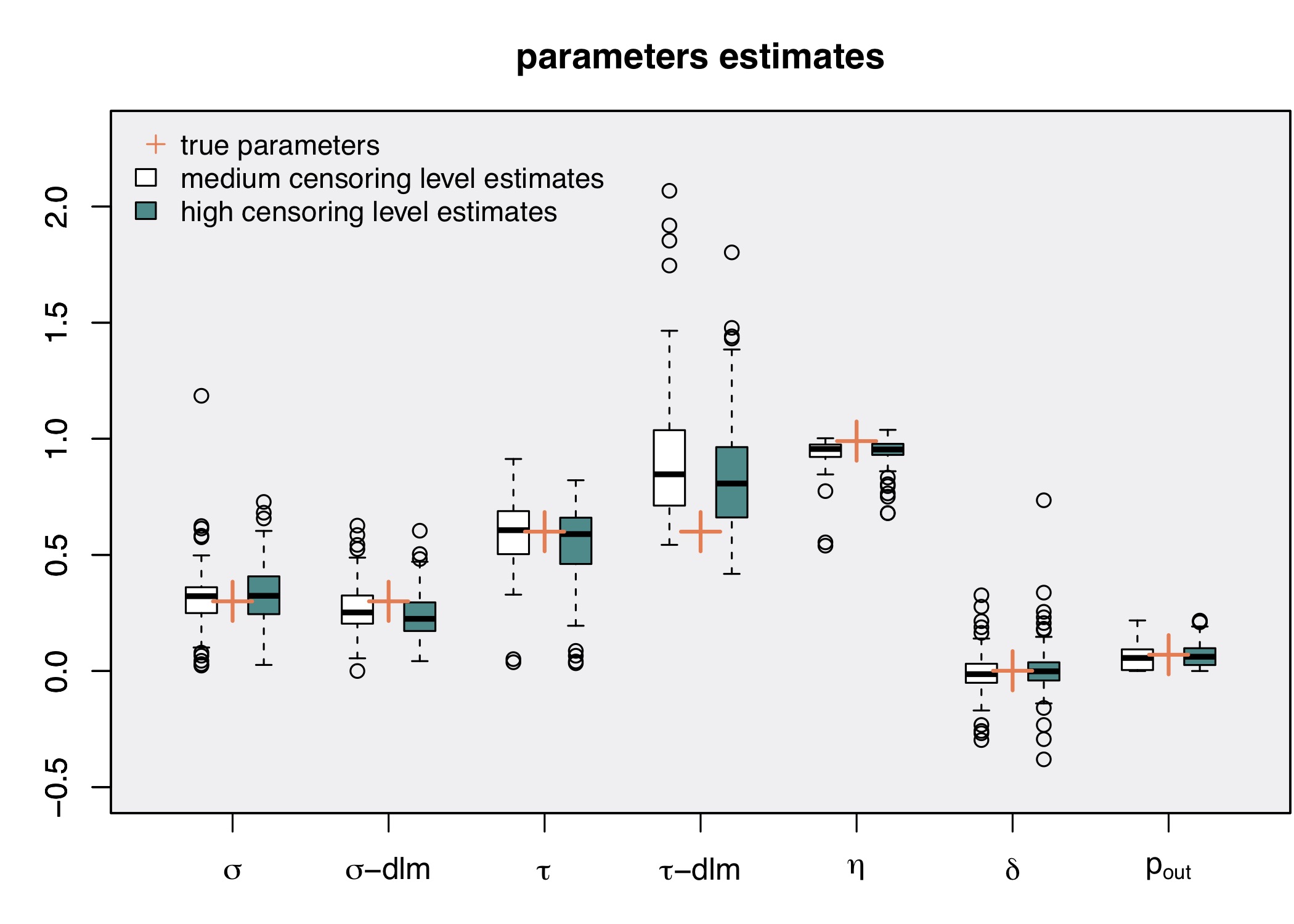}
\caption{\label{fig:param} Parameters estimates with our method ($\sigma$, $\tau$,$\eta$, $\delta$) and with a 2-parameter Kalman Smoother ($\sigma$-dlm and $\tau$-dlm) for $100$ replicates of the simulation experiment ($150$ time steps, outlier rate of \new{$p=7\%$}, 
observation rate of $50\%$), for $16\%$ and $31\%$ of censored data (resp. white and cyan boxes).}
\end{figure}

Figure~\ref{fig:param} shows that the parameters are correctly estimated with the proposed method. However, one can notice a little negative bias in $\eta$ estimates which might be due to the asymmetric impact of the estimation bias, with a strong degradation if the $\eta$ estimate exceeds one. 
The parameters learned by our method are indeed close on average to the parameters used for the simulation of the data. The parameters learned with the 2-parameter Kalman smoother ($\sigma$-dlm and $\tau$-dlm) present an estimation bias which illustrates the necessity to take into account the censoring of the data and the outliers when they exist. 

\subsubsection{Outliers identification}
The \emph{a posteriori} probability that an observation $Y_t$ is an outlier, $\mathbb{P}(O_t=1|\newn{Y_{\mathcal{T}}=y_{\mathcal{T}}})$, is computed as specified Equation~\ref{eq:outliers} with, for each simulation, the estimated parameters. 
The ROC (Receiver Operating Characteristic) curves for the detection of the simulated outliers (around $525$ out of about $\frac{n\times 100}{2}=7500$ observations of $Y$), across all simulations taken altogether, 
show \new{correct} 
outlier detection performances despite the censoring of some outliers and the possibility for the outliers to take values very close to $X$s (the outliers being simulated from a Uniform distribution). 
Indeed, the AUC (Area Under the Curve) 
are of \new{$0.74 \pm 0.01$ for both censoring levels}.  
When the true outlier a priori probability, $p=0.07$, is provided, they are respectively of $0.817 \pm 0.014$ and $0.767 \pm 0.016$ for the medium and high censoring settings. 
The standard deviation for the computed AUC are here obtained by discounted sampling the $7500$ available observations $1000$ times.


\subsubsection{Prediction of the underlying process distribution}

\begin{figure}[h!]
    \centering
    \includegraphics[scale=0.2]{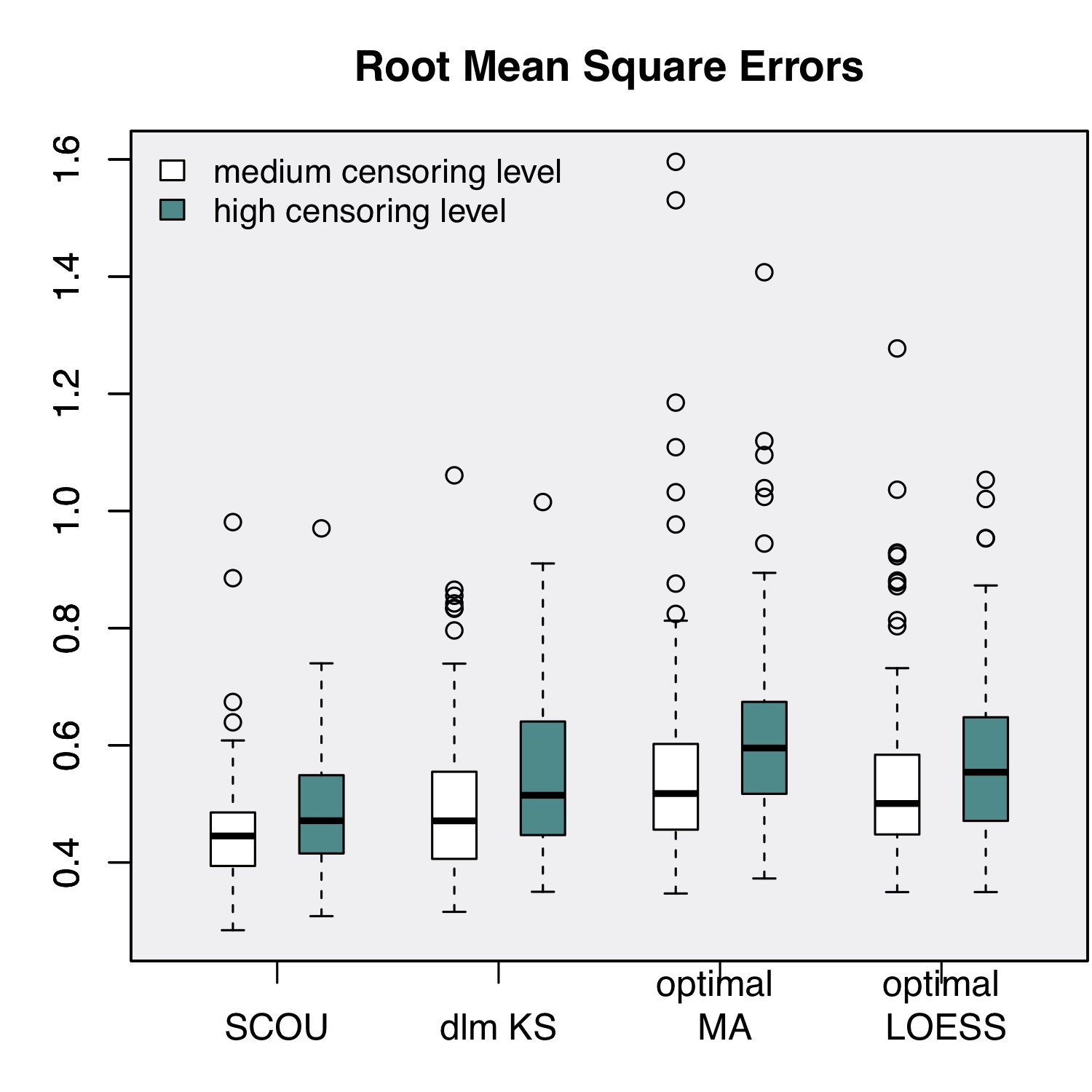}
    \caption{\label{fig:rmse} Root Mean Squared Errors for the prediction of $X$ \newnnn{for 4 competitive smoothing methods (the proposed smoother (SCOU), the 2-parameter Kalman Smoother, the LOESS and the moving average (MA))}, with $16\%$ and $31\%$ of censored data (resp. white and cyan boxes).}
\end{figure}

\begin{figure}[h!]
    \centering
    \includegraphics[scale=0.2]{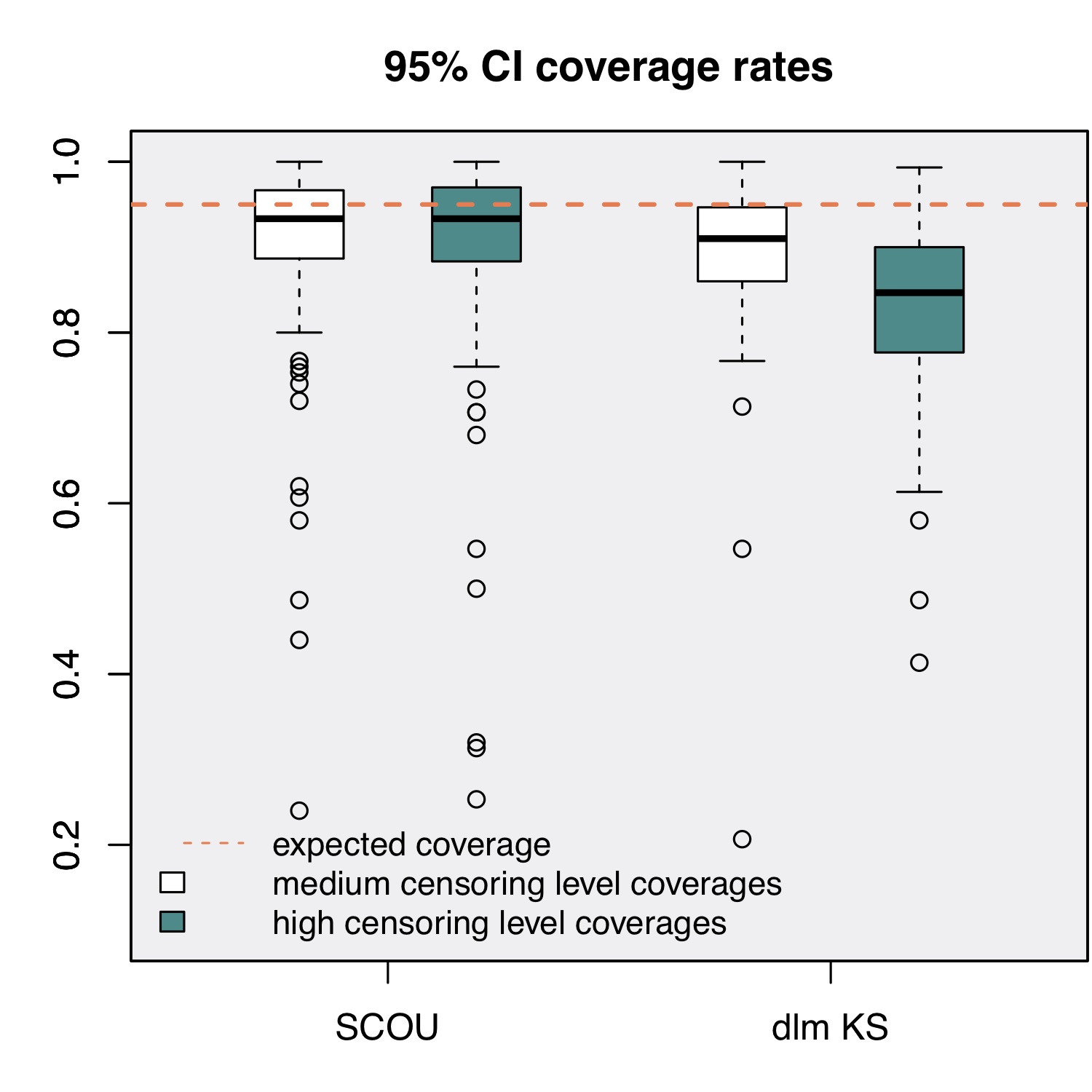}
    \caption{\label{fig:txcov} Coverage rates of the prediction intervals for the prediction of $X$, with $16\%$ and $31\%$ of censored data (resp. white and cyan boxes).}
\end{figure}

Finally, we evaluate the ability of our method to predict the distribution of $X$ conditionally on the set of observed $Y$, i.e. the distribution of  $(X|Y_\mathcal{T}=y_\mathcal{T})$ .  
As illustrated Figure~\ref{fig:rmse}, in both the medium and high censoring level experimental settings, the 
RMSEs obtained by our method 
are significantly lower than the ones obtained 
with the 2-parameter Kalman Smoother, 
with the LOESS method and with the moving average 
\new{even when their parameters (respectively the span and the number of days) are chosen so as to minimize the RMSE obtained by leave-one-out cross validation}.\\
As for the variance prediction, the coverage rates of the $95\%$ Prediction Intervals of our method, derived from the predicted distributions of $(X_t|\newn{Y_{\mathcal{T}}=y_{\mathcal{T}}})$, are (on average) close to the target of $95\%$, with a median coverage rate of \new{$93\%$} (compared to \new{$91\%$} with the 2-parameter Kalman Smoother) in the medium censoring level setting (resp. \new{$93\%$} and \new{$85\%$} in the high censoring level setting) as illustrated Figure~\ref{fig:txcov}.

\section{Application to the data from \emph{Obépine}}
\label{sec:appli}

The developed smoothing method 
aims to provide an estimate of the actual amount of viral genome arriving at each WWTP and to assess the uncertainty of this estimation. 

\new{The concentration measurements provided by \emph{Obépine} are adjusted beforehand} for rainfalls and wastewater sources other than from households, which can otherwise distort the conclusions when it is abundant by diluting the water arriving at the WWTPs  \newnnn{\citep{cluzel2022nationwide}}. 

The direct application of \new{SCOU} on the flow-adjusted measurements shows heteroscedasticity of the residuals (estimated by $y_\mathcal{T}-\mathbb{E}(X_\mathcal{T}|Y_\mathcal{T}=y_\mathcal{T})$): their variance increases with $y$ (even after removal of numerous measurements identified as outliers and of censored measurements).
Besides, the underlying process, $X$, follows the dynamic of an epidemic. During the exponential growth, it is thus supposed to multiply from one day to another. 
For both those reasons, a logarithmic transformation of the \emph{Obépine} data was performed prior to applying the proposed method. 

Hence, for this data set, 
$Y$ is the logarithm of the \new{flow-adjusted} measured concentrations, 
$X$ is the logarithm of the actual virus concentrations (to be estimated), 
$\ell$ is the logarithm of the limit below which censoring occurs (typically $\log(1000U/L)$ for the quantification of the SARS-CoV-2 E gene, where U stands for RNA Units, \new{but it can fluctuate from one day to another due to flow adjustments}). 

\begin{figure}[h!]
\centering
\includegraphics[scale=0.22]{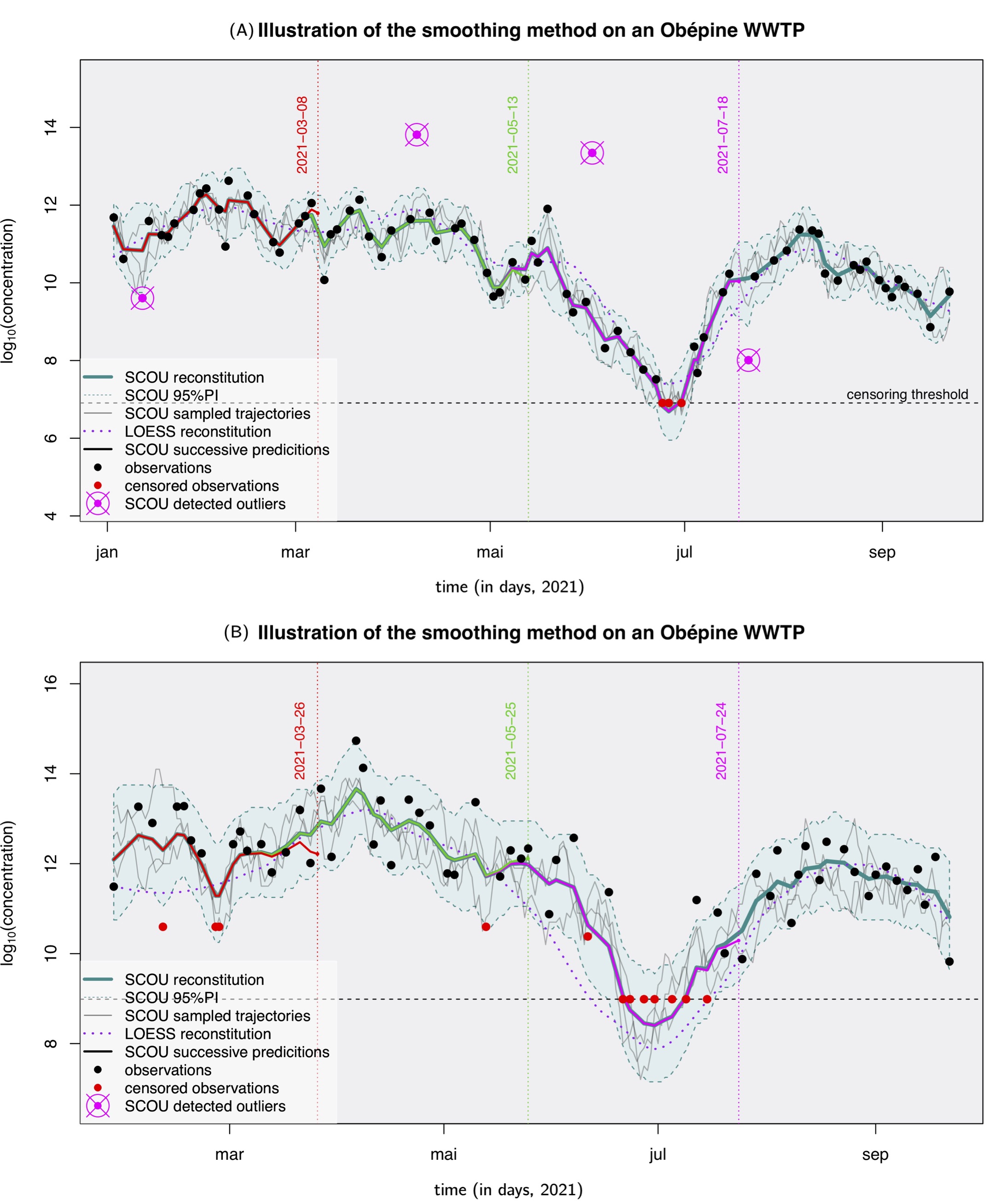}
\caption{\label{fig:datareelles} 
\textbf{(A)} and \textbf{(B)}:  application of \new{the LOESS (for a span parameters chosen by leave-one-out) and of} the proposed smoother for two WWTPs of the \emph{Obépine} network:  \newn{examples of successive predictions} for the (never observed) underlying signal, $X$,  \newn{with the final parameters}, $3$ simulations of $X$ according to the proposed model, $95\%$ prediction intervals for $X$ and outlier detection results.} 
\end{figure}

The application of \new{SCOU} to real data from the \emph{Obépine} network is illustrated for two WWTPs on Figure~\ref{fig:datareelles}. \newn{As shown by the successive predictions, once the parameters are fixed, the predictions are rather stable from one day to another.} 

\subsection{Parameters estimates for the \emph{Obépine} data}

\begin{figure}[h!]
\centering
\includegraphics[scale=0.5]{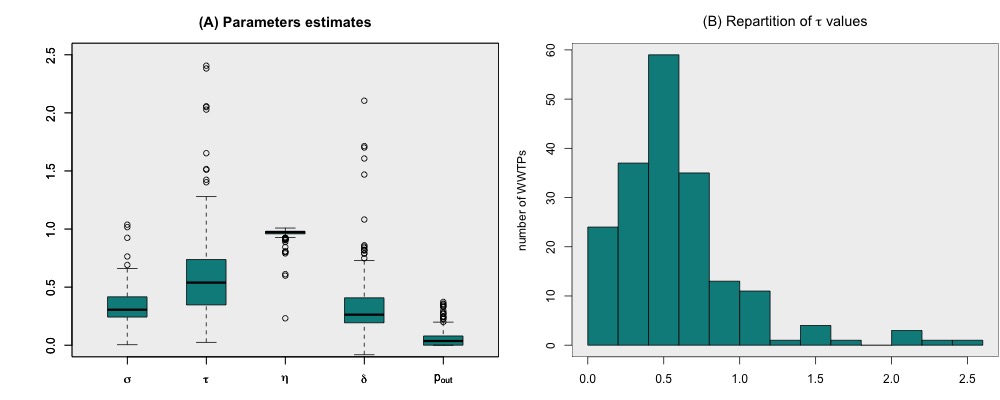}
\caption{\label{fig:epsirep} 
\textbf{(A)} Parameters estimates for \newn{$190$} \emph{Obépine} WWTPs and \textbf{(B)} detailed distribution of the corresponding $\tau$ estimates which gives an estimation of the error of the whole measurement system.} 
\end{figure}

The estimates of the smoothing parameters for the \newn{$190$} \emph{Obépine} WWTPs \newn{with enough available observations (at least $10$ measurements)} 
are illustrated in panel \textbf{(A)} of Figure~\ref{fig:epsirep}. 
\newnn{The estimated $\delta$ parameters are above $0$ while $\eta$ parameters are very close to $1$. Hence, when no new information is provided (no new $Y$ is observed), the smoother predicts, in average, an increase in $X$ values.} 

The uncertainty of the monitoring system is evaluated by the parameter $\tau$, whose estimates distribution for the \newn{$190$} \emph{Obépine} WWTPs is illustrated in panel \textbf{(B)}  of Figure~\ref{fig:epsirep}. 
\new{The average value for $\tau$ is $ \newn{0.54} \log(U/L)$.} 
This corresponds to a standard deviation of about $\newn{0. 65 x}$, 
where $x$ is the real E-gene \newnn{RNA} concentration (in $U/L$) to which the standard deviation of the measurement error is supposed proportional in our model. The allocation of this uncertainty between the sampling error, the RT-qPCR \newnn{or RT-dPCR} error and other possible sources of error throughout the system is however not known.  

\begin{figure}[h!]
\centering
\includegraphics[scale=0.15]{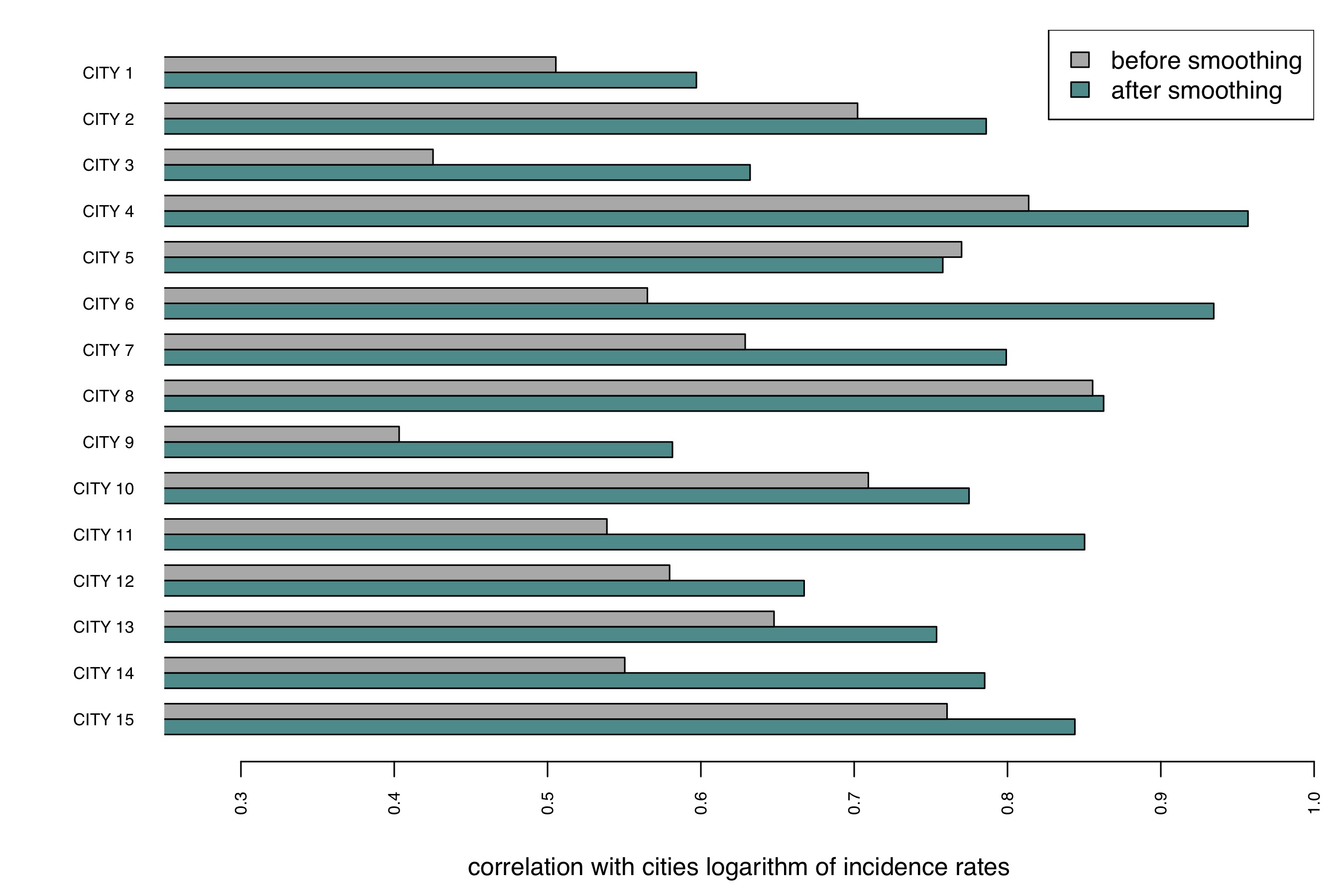}
\caption{\label{fig:corincid} Correlation of raw \emph{Obépine} data and flow-adjusted and smoothed \emph{Obépine} data 
with the logarithm of incidence rates of the corresponding cities for $15$ cities with enough data available on both sides, for a lag time ranging from $1$ to $3$ days in both directions and for a period ranging from the beginning of the second wave to the $21^{st}$ of May 2021.}
\end{figure}

Importantly, the resulting \new{smoothed signal}
is well correlated with the logarithm of the local COVID-19 incidence rates 
and this correlation is most of the time greatly enhanced by the proposed smoothing step as depicted Figure~\ref{fig:corincid} for the $15$ cities with enough data available on both sides. 
On this figure, the correlations are only computed for dates at which raw data was available and corresponds to the best correlation for a time lag ranging from 1 to 3 days in one direction or the other. They are computed for a period ranging from the beginning of the second wave to the $21^{st}$ of May 2021. 
Those correlations are not expected to be higher since, contrary to incidence rates, the indicator also points to asymptomatic peoples infected by SARS-CoV-2 and is not biased by people getting tested outside their city (during holidays for instance) nor by varying testing policies.

In order to produce comparable values from one WWTP to another, \emph{Obépine} moreover performs a scaling of the data after smoothing  \newnnn{\citep{cluzel2022nationwide}}. This scaling takes into account, for example, the maximum amount of virus measured during the \newnn{epidemic wave that occurred in Autumn 2020 in France}, the range of volume treated by the WWTP and the specifics of the laboratory that analyzed the sample by RT-qPCR  \newnn{or RT-dPCR}. 
 \new{The final computed indicator (called WWI for WasteWater Indicator) shows good behavior with regard to the corresponding incidence rates \citep{cluzel2022nationwide}}.

\section{Discussion}

We developed a method to smooth one dimensional time-series consisting of successive censored measurements with outliers 
when the associated measurement uncertainty is not known and \newn{the measured quantities} have an auto-regressive nature. By discretizing the state space of the monitored quantities, the proposed method has the advantage of being easily adaptable to \newn{the} specificities of the data (such as measurement censoring and the occurrence of outliers). 
An experiment on artificial data validates the proposed inference and prediction method. 
Our method has then been successfully applied to data generated during the \emph{Obépine} monitoring of SARS-CoV-2 genome concentration in WWTPs  \newnnn{\citep{cluzel2022nationwide}}. Importantly,
the proposed smoothing procedure enhances the correlation of the data with other epidemiological indicators such as the incidence rate of COVID-19. The time lag between the two signals is moreover just a few days  \citep{cluzel2022nationwide}, making the WWI a credible alternative to 
the evaluation of the incidence rate 
through massive individual testing. This approach may be especially relevant if massive testing campaigns become less relevant notably with the advancement of the vaccination campaign and the availability of self-tests to the general public. 
Both of these factors may \newn{indeed} induce a progressive but significant decline in participation in testing in a few months and a significant dwindling of the population surveyed to monitor the pandemic, potentially making it even more partial than it is now.

The proposed method could be further developed. 
First, the underlying $X$ process could have a longer time dependency than an AR(1) process. We could thus develop an AR(p) version of this method to handle this (with $p>1$).
Besides, the behavior of the marginal $X$ process, and thus its parameters, 
are expected to change as we move from a propagation of the epidemic stage to a decreasing stage \citep{berestycki2021effects}. 
Joint treatment of the WWTPs time-series could overcome the lack of individual data to face this problem. 
One could, for example, automatically detect common breakpoints corresponding to a change in the parameters $\eta$ or $\delta$. Another possibility would be either to use extrinsic knowledge of the reproduction factor of the epidemic as input data or to add it as a latent variable that slowly evolves from one day to another.\\
Another way to proceed would be to deduce from other available epidemiological data the shape of the signal to be found in wastewater (and thus an adequate smoothing) based on a fine modeling of the whole pathway of SARS-CoV-2 from the human population to wastewater such as the one proposed by \cite{nourbakhsh2021wastewater}. However, such a mechanistic representation includes a large number of unknowns (actual number of infected individuals in the population, rate of RNA degradation in wastewater,...) which makes it difficult to exploit for the reconstruction purposes aimed here.

\section*{Conflict of Interest Statement}

The authors declare that the research was conducted in the absence of any commercial or financial relationships that could be construed as a potential conflict of interest.

\section*{Author Contributions}

Yvon Maday, Jean-Marie Mouchel, Vincent Maréchal, Laurent Moulin and Sébastien Wurtzer brought on the scientific problem. 
Marie Courbariaux contributed to the design of the algorithm, performed experiments on artificial and real data, and wrote the paper.
Grégory Nuel contributed to the design of the algorithm and coordinated the experiments and the writing of the paper. 
Nicolas Cluzel and Siyun Wang prepared the data provided by \emph{Obépine},  
\new{contributed to the design of the algorithm, performed experiments on artificial and real data} 
and discussed the results. 
Nicolas Cluzel, Siyun Wang, Jean-Marie Mouchel, Yvon Maday and Vincent Maréchal proofread and discussed the content of the paper. 
Yvon Maday moreover coordinated the exchanges about the article within the \emph{Obépine} consortium.

\section*{Funding}
This work was carried out within the \emph{Obépine} 
project funded by the French Minister of Higher Education, Research and Innovation (Ministre de l'Enseignement Supérieur, de la Recherche et de l'Innovation). Financial support was also obtained from the French National Center for Scientific Research and Sorbonne Université.

\section*{Acknowledgments}
We would like to thank the WWTPs and the laboratories that contribute day to day to the \emph{Obépine} network.
In particular, the LCPME (Isabelle Bertrand and Christophe Gantzer) produced three quarters of the raw data that was used to produce Figure~\ref{fig:corincid}. 
The experimental 
reflections of this work were fed by exchanges with the members of the \emph{Obépine} consortium and notably Karine Laurent. 

\newnn{
\section*{The \emph{Obépine} consortium}
The \emph{Obépine} consortium's Scientific Coordination and Steering Committee (CCOS) is composed of Isabelle Bertrand, Mickael Boni, Christophe Gantzer, Soizick F. Le Guyader,  Yvon Maday, Vincent Maréchal, Jean-Marie Mouchel, Laurent Moulin, Rémy Teyssou and Sébastien Wurtzer.
The Scientific Interest Grouping (GIS) \emph{Obépine} is directed by Vincent Maréchal associated with two co-directors: Christophe Gantzer and Laurent Moulin. It gathers the CNRS, Eau de Paris, EPHE, Ifremer, Inserm, IRBA, Sorbonne University, Clermont Auvergne University, Lorraine  University and Université de Paris.
}


\section*{Data Availability Statement}
\newnnn{WWI produced by Obépine are freely available for all WWTPs treated by the \emph{Obépine} network on the data.gouv.fr public platform. Incidence rate data are partially available in open access for 22 Public Establishments of Intermunicipal Cooperation (EPCI) and can be found on the same platform.}

\bibliographystyle{plainnat}
\bibliography{biblio}

\end{document}